\documentclass[prb,twocolumn,floatfix,showpacs]{revtex4}
\pdfoutput=1
\usepackage{graphicx}
\usepackage{amsmath}
\usepackage{bm}
\begin{document}

\title{Topological quantum computation away from the ground state with Majorana fermions}
\author{A. R. Akhmerov}
\affiliation{Instituut-Lorentz, Universiteit Leiden, P.O. Box 9506, 2300 RA Leiden, The Netherlands}
\date{May 2010}
\begin{abstract}
We relax one of the requirements for topological quantum computation with Majorana fermions. Topological quantum computation was discussed so far as manipulation of the wave function within degenerate many body ground state. The simplest particles providing degenerate ground state, Majorana fermions, often coexist with extremely low energy excitations, so keeping the system in the ground state may be hard. We show that the topological protection extends to the excited states, as long as the Majorana fermions do not interact neither directly, nor via the excited states. This protection relies on the fermion parity conservation, and so it is generic to any implementation of Majorana fermions.
\end{abstract}
\pacs{03.67.Lx, 03.67.Pp, 71.10.Pm, 74.90.+n}
\maketitle

Topological quantum computation is manipulation of the wave function within a degenerate many-body ground state of many nonabelian anyons. Interchanging the anyons applies a unitary transformation to the ground state wave function. The simplest of the nonabelian anyons useful for topological quantum computation are Majorana fermions. These are expected to exist in 5/2 fractional quantum Hall effect \cite{moore_nonabelions_1991} and in certain exotic superconductors \cite{volovik_fermion_1999,read_paired_2000,fu_superconducting_2008,sau_generic_2010}. In 5/2 fractional quantum Hall effect the Majorana fermions are charge e/4 quasiholes, and in superconductors Majorana fermions are zero energy single particle states either trapped in vortex cores or other inhomogeneities.\cite{volovik_fermion_1999,kitaev_unpaired_2001,fu_josephson_2009,wimmer_majorana_2010} 

Superconducting implementations of Majorana fermions potentially allow for a larger bulk gap of a few Kelvin as compared with 500 mK for fractional quantum Hall effect. One significant difference between the superconductors and the fractional quantum Hall effect is that Majorana fermions in superconductors appear where the superconducting gap in excitation spectrum closes. This means that Majorana fermions would not be isolated from other excitations by the bulk gap, but coexisting with a lot of bound fermionic states with level spacing of the order of the minigap $\Delta^2/E_F$, where $\Delta \sim 1$ K is the superconducting gap and $E_F$ the fermi energy.\cite{caroli_bound_1964} If $E_F\sim 1$ eV, minigap is at least a thousand times smaller than the bulk gap, so coupling between Majorana states and excited states is unavoidable with existing experimental methods. Already detection of Majorana fermions becomes problematic in this regime and requires ballistic samples and spatial resolution of density of states on the scale of Fermi wave length.\cite{kraus_testing_2008} This is why there is research aimed at increasing the minigap.\cite{sau_robustness_2009}

We adopt a different strategy and show that coupling to excited states does not remove the topological protection as long as different Majorana fermions stay decoupled. The topological protection persists because coupling to excited states has to preserve the global fermion parity. Using only the conservation of the global fermion parity and the fact that different Majorana fermions are well separated we identify new Majorana operators, which are protected even if the original Majorana fermions coexist with many excited states. We also check that the braiding rules for the new Majorana operators are the same as for original ones.

We start from a brief introduction to Majorana fermions, for more information see e.g. Ref.~\onlinecite{nayak_non-abelian_2008}. A single Majorana fermion is described by a fermionic annihilation operator $\gamma$ which is equal to the creation operator
\begin{equation}
 \gamma=\gamma^\dagger.\label{eq:majoranadefine}
\end{equation}
Due to this defining property of Majorana fermions they are also called ``real fermions'' or ``particles equal to their own antiparticles''. Substituting Eq.~\ref{eq:majoranadefine} into the fermion anticommutation relation we get
\begin{equation}
 \{\gamma,\gamma^\dagger\}=2\gamma^2=2\gamma^\dagger\gamma=1.
\end{equation}
The last equality is a manifestation of the fact that a single Majorana fermion is pinned to the fermi level and accordingly is always half-filled. Additionally it is not possible to add a perturbation to the Hamiltonian, which would move a single Majorana level away from fermi level, at least two Majorana fermions are required. The only possible coupling term between two Majorana fermions has the form
\begin{equation}
H_\gamma = i \varepsilon \gamma_1 \gamma_2.
\end{equation}
The perturbation $H_\gamma$ hybridizes two Majorana states into a single complex fermion state at energy $\varepsilon$ and with creation and annihilation operators
\begin{equation}
a^\dagger_{12} = \frac{\gamma_1 + i \gamma_2}{\sqrt{2}},\text{\;\;}a_{12} = \frac{\gamma_1 - i \gamma_2}{\sqrt{2}}. \label{eq:complexfermion}
\end{equation}
If Majorana fermions are well separated, the coupling between them decays exponentially with the distance between them.\cite{read_paired_2000,kraus_testing_2008} Additionally if the superconductor is grounded, the charging energy also vanishes, leaving the Majorana fermions completely decoupled.\cite{fu_electron_2010} In the limit when coupling between Majorana fermions $\varepsilon$ is negligibly small, $H_\gamma$ has two zero energy eigenstates which differ by fermion parity
\begin{equation}
 (1-2 a^\dagger_{12} a_{12})= 2 i \gamma_1 \gamma_2.
\end{equation}
If the system has $N$ decoupled Majorana fermions, the ground state has $2^{N/2}$ degeneracy and it is spanned by fermionic operators with the form \eqref{eq:complexfermion}. Braiding Majorana fermions performs unitary rotations in the ground state space and makes the basis for topological quantum computation.

 To understand how coupling with excited states gives nontrivial evolution to the wave function of Majorana fermions we begin from a simple example. We consider a toy model containing only two Majorana fermions $\gamma_1$ and $\gamma_2$ and a complex fermion $a$ bound in the same vortex as $\gamma_1$. At $t=0$ we turn on the coupling between $\gamma_1$ and $a$ with Hamiltonian
\begin{equation}
 H_{a1}=i \varepsilon (a+a^\dagger)\gamma_1.
\end{equation}
At $t=\pi\hbar/\varepsilon$ we turn off $H_{a1}$ and give finite energy to the fermion by a term $\varepsilon a^\dagger a$. We denote by $|0\rangle$ the state where two Majorana fermions share no fermion, so an eigenstate of $2 i \gamma_1 \gamma_2$ with eigenvalue $1$, and by $|1\rangle$ the eigenstate of $2 i \gamma_1 \gamma_2$ with eigenvalue $-1$. If the system begins from a state $|0\rangle$, then it evolves into an excited state $a^\dagger|1\rangle$, so the Majorana qubit flips. This seems to destroy the topological protection, however there is one interesting detail: since there are two degenerate ground states $|0\rangle$ and $|1\rangle$, there are also two degenerate excited states: $a^\dagger|0\rangle$ and $a^\dagger|1\rangle$. So while $|0\rangle$ changes into $a^\dagger|1\rangle$, $|1\rangle$ changes into $a^\dagger|0\rangle$. The two end states differ by total fermion parity, which is the actual topologically protected quantity. In the following we identify the degrees of freedom which are protected by nonlocality of Majorana fermions and do not rely on the system staying in the ground state.

 We consider a system with $N$ vortices or other defects trapping Majorana fermions with operators $\gamma_i$, where $i$ is the number of the vortex. Additionally every vortex has a set of $m_i$ excited complex fermion states with creation operators $a_{ij}$, with $j\le m_i$ the number of the excited state. We first consider the excitation spectrum of the system when the vortices are not moving and show that it is possible to define new Majorana operators which are protected by fermion parity conservation even when there are additional fermions in the vortex cores. Parity of all the Majorana fermions is given by $(2 i)^{n/2}\prod_{i=1}^N\gamma_i$, so the total fermion parity of $N$ vortices, which is a fundamentally preserved quantity, is then equal to
\begin{multline}
 {\cal P} = (2 i)^{n/2}\prod\limits_{i=1}^N\gamma_i \times\prod\limits_{i=1}^N \prod\limits_{j=1}^{m_i} [1-2 a_{ij}^\dagger a_{ij}] \\= (2 i)^{n/2}\prod\limits_{i=1}^N \left(\prod\limits_{j=1}^{m_i} [1-2 a_{ij}^\dagger a_{ij}]\gamma_i\right).
\end{multline}
This form of parity operator suggests to introduce new Majorana operators according to
\begin{equation}
 \Gamma_i=\prod\limits_{j=1}^{m_i} [1-2 a_{ij}^\dagger a_{ij}]\gamma_i.
\end{equation}
It is easy to verify that $\Gamma_i$ satisfy the fermionic anticommutation relations and the Majorana reality condition \eqref{eq:majoranadefine}.
The total fermion parity written in terms of $\Gamma_i$ mimics the fermion parity without excited states in the vortices
\begin{equation}
 {\cal P} = (2 i)^{n/2} \prod\limits_{i=1}^N \Gamma_i,
\end{equation}
so the operators $(2i)^{1/2}\Gamma_i$ can be identified as \emph{the local part of the fermion parity operator} belonging to a single vortex.
We now show that the operators $\Gamma_i$ are protected from local perturbations. Let the evolution of system be described by evolution operator
\begin{equation}
 U= U_1\otimes U_2\otimes \cdots\otimes U_n, 
\end{equation}
with $U_i$ evolution operators in $i$-th vortex. The system evolution must necessarily preserve the full fermion parity
\begin{equation}
 {\cal P}=U^\dagger{\cal P}U,
\end{equation}
and hence

\begin{multline}
(2 i)^{n/2}\prod\limits_{i=1}^N \Gamma_i=(2 i)^{n/2} \prod\limits_{i=1}^N U_i^\dagger  \times \prod\limits_{i=1}^N \Gamma_i  \\ \times \prod\limits_{i=1}^N U_i=
(2 i)^{n/2}\prod\limits_{i=1}^N U_i^\dagger \Gamma_i U_i.
\end{multline}
This equation should hold for any set of allowed $U_i$. Taking $U_i=\bm{1}$ for all $i \neq j$ we come to
\begin{equation}
 U_j^\dagger \Gamma_j U_j = \Gamma_j,\label{eq:invariance}
\end{equation}
for any $U_j$. In other words, the new Majorana operators $\Gamma_j$ are indeed not changed by any possible local perturbations.

We now need to show that the protected Majorana operators $\Gamma_i$ follow the same braiding rules \cite{ivanov_non-abelian_2001} as the original ones. The abelian part of braiding, namely the Berry phase,\cite{stern_geometric_2004,read_non-abelian_2009} is not protected from inelastic scattering in vortices, so it will be completely washed out. The non-abelian part of the braiding rules is completely described by the action of the elementary exchange of two neighboring vortices $T$ on the Majorana operators. As shown in Ref.~\onlinecite{ivanov_non-abelian_2001}, exchanging Majorana fermions $\gamma_i$ and $\gamma_j$ is described by $\gamma_i\rightarrow \gamma_j$, $\gamma_j \rightarrow -\gamma_i$. The fermion parity operators $(1-2 a^\dagger_{ij}a_{ij})$ have trivial exchange statistics as any number operators. 
Applying these rules to exchange of two vortices containing excited states gives
\begin{subequations}
\begin{multline}
\Gamma_i=\prod\limits_{k=1}^{m_i} [1-2 a_{ik}^\dagger a_{ik}]\gamma_i \rightarrow
\prod\limits_{k=1}^{m_j} [1-2 a_{jk}^\dagger a_{jk}]\gamma_j= \Gamma_j,
\end{multline}
\begin{multline}
\Gamma_j=\prod\limits_{k=1}^{m_j} [1-2 a_{jk}^\dagger a_{jk}]\gamma_j \rightarrow\\
\prod\limits_{k=1}^{m_i} [1-2 a_{ik}^\dagger a_{ik}](-\gamma_i)= -\Gamma_i.
\end{multline}
\end{subequations}
This finishes the proof that braiding rules are the same for $\Gamma_i$.

Our proof of protection of Majorana fermions and their braiding properties from conservation of fermion parity only relies on particle statistics of Majorana and complex fermions. Consequently it fully applies to the Moore-Read state of 5/2 fractional quantum Hall effect, p-wave superfluids of cold atoms,\cite{gurarie_resonantly_2007} or any other implementation of Majorana fermions. Part of this proof can be reproduced using topological considerations in the following manner. If a perturbation is added to the Hamiltonian and additional excitations are created in a vortex, the fusion outcome of all these excitations cannot change unless these excitations are braided or interchanged with those from other vortices. So if a system is prepared in a certain state, then excitations are created in vortices, braiding is performed and finally the excitations are removed, the result has to be the same as if there were no excitations. Our proof using parity conservation, however, allows additionally to identify which part of the Hilbert space stays protected when excitations are present. Since removing the low energy excitations does not seem feasible, this identification is very important. It allows a more detailed analysis of particular implementations of the quantum computation with Majorana fermions. For example we conclude that implementation of the phase gate using charging energy, as described in Ref.~\onlinecite{hassler_anyonic_2010}, does not suffer from temperature being larger than the minigap since it relies on fermion parity, not on the wave function structure.

Since all the existing readout schemes of a Majorana qubit\cite{kitaev_fault-tolerant_2003,fu_superconducting_2008,fu_probing_2009,akhmerov_electrically_2009,grosfeld_aharonov-casher_2010} are measuring the full fermion parity of two vortices, and not just the parity of the fermion shared by two Majorana fermions, all these methods also work if Majorana fermions coexist with excited states. The signal strength however is reduced significantly when the temperature is comparable with the minigap due to dephasing of the internal degrees of freedom of vortices. Using interferometry of Josephson vortices\cite{hassler_anyonic_2010}, which do not trap low energy excitations allows to avoid this problem.

In conclusion, we have shown that topological quantum computation with Majorana fermions is not sensitive to presence of additional localized states coexisting with Majorana fermions in superconducting vortices. This significantly relaxes the requirements on the temperature needed to achieve topological protection of Majorana fermions. 

I have benefited from discussions with J. K. Asboth, J. H. Bardarson, C. W. J. Beenakker, L. Fu, F. Hassler, C.-Y. Hou, and  A. Vishwanath. This research was supported by the Dutch Science Foundation NWO/FOM.

\bibliography{finite_energy_tqc}

\end{document}